\documentclass[onecolumn]{pasj01}
\SetRunningHead{Hosono et al.}{Unconvergence of Very Large Scale GI Simulations}

\Received{2016/12/09}
\Accepted{2016/12/21}


\maketitle

\begin{document}

\title{Unconvergence of Very Large Scale GI Simulations}

\author{Natsuki \textsc{Hosono}\altaffilmark{1,2}}
\author{Masaki \textsc{Iwasawa},\altaffilmark{2}}
\author{Ataru \textsc{Tanikawa}\altaffilmark{3,2}}
\author{Keigo \textsc{Nitadori}\altaffilmark{2}}
\author{Takayuki \textsc{Muranushi}\altaffilmark{2}}
\author{Junichiro \textsc{Makino}\altaffilmark{4,2,5}}

\altaffiltext{1}{Graduate School of Advanced Integrated Studies in Human Survivability, Kyoto University, 1 Nakaadachi-cho, Yoshida, Sakyo-ku, Kyoto 606-8306, Japan}
\altaffiltext{2}{RIKEN Advanced Institute for Computational Science, 7-1-26 Minatojima-minami-machi, Chuo-ku, Kobe, Hyogo, Japan}
\altaffiltext{3}{Department of Earth and Astronomy, College of Arts and Science, The University of Tokyo, 3-8-1 Komaba, Meguro-ku, Tokyo, Japan}
\altaffiltext{4}{Department of Planetology, Graduate School of Science Faculty of Science, Kobe University, 1-1, Rokkodai-cho, Nada-ku, Kobe, Hyogo 657-8501, Japan}
\altaffiltext{5}{Earth-Life Science Institute, Tokyo Institute of Technology, 2-12-1 Ookayama, Meguro-ku, Tokyo, Japan}

\email{hosono.natsuki.2a@kyoto-u.ac.jp}

\begin{abstract}
The giant impact (GI) is one of the most important hypotheses both in planetary science and geoscience, since it is related to the origin of the Moon and also the initial condition of the Earth.
A number of numerical simulations have been done using the smoothed particle hydrodynamics (SPH) method.
However, GI hypothesis is currently in a crisis.
The ``canonical'' GI scenario failed to explain the identical isotope ratio between the Earth and the Moon.
On the other hand, little has been known about the reliability of the result of GI simulations.
In this paper, we discuss the effect of the resolution on the results of the GI simulations by varying the number of particles from $3\times10^3$ to $10^8$.
We found that the results does not converge, but shows oscillatory behaviour.
We discuss the origin of this oscillatory behaviour.
\end{abstract}


\section{Introduction}\label{sec:introduction}
The giant impact (GI) hypothesis plays an important role both in planetary science and geoscience.
It was first suggested as the origin of the Moon by \citet{HD75} and \citet{CW76}.
According to the GI scenario, a Mars-sized impactor hits the proto-Earth and generates substantial amounts of the debris disc around the proto-Earth, which later accumulates into the Earth's Moon.
Since this scenario can explain characteristic features of the current Earth-Moon system such as the large angular momentum and the small core of the Moon, it has become widely accepted as the standard scenario for the formation of the Moon.
So far, a number of numerical simulations for this scenario have been carried out mainly by the smoothed particle hydrodynamics (SPH; \cite{L77, GM77}) method (e.g., \cite{B+86, CA01, NS14}).

However, the GI scenario is currently in a crisis; recent high-precision isotope ratio measurement revealed that the isotope ratios of the bulk of the Moon in several elements are very close to that of the Earth (e.g., \cite{W+01, T+07, G+07, Z+12}).
The proto-Earth and the impactor are likely to have different isotope ratios because the isotope ratio of an object reflects the place where the object was formed.
A simple explanation for this close agreement of the isotope ratios is that the Moon was formed from a proto-Earth dominated disc.
Previous studies based on numerical simulations, however, concluded that the so-called ``canonical'' GI scenarios would result in impactor-dominated discs.
This discrepancy should be resolved.

Lately, several re-investigations of numerical simulations of GI have been reported.
Simulations with methods other than the standard SPH method have been done, e.g., static mesh \citep{W+06}, adaptive mesh \citep{C+13} and Density Independent SPH (DISPH; \cite{H+16}).
\citet{W+06} concluded that the results with static mesh code were significantly different from those with the standard SPH.
They, however, adopted much simpler equation of state to represent the bodies.
Thus, it is not straightforward to compare their results to those of other works.
\citet{C+13} concluded that the results with adaptive mesh give similar results to those with the standard SPH.
\citet{H+16} performed the numerical simulations of GI using DISPH, an improved formulation of SPH, and concluded that DISPH tends to form more compact disc than the standard SPH does.

Resolution is also important for the validity of numerical simulations.
Thus, it is very important to test the convergence in terms of the resolution, namely, the dependence of the results on the number of SPH particles used in a simulation.
We expect the numerical solutions to converge to the physical solution, in the limit of the infinite number of particles.
\citet{C+13} varied the number of particles from $10^4$ to $10^6$ and compared the results.
They concluded that the predicted moon masses obtained from each simulation are similar.
However, if we closely investigate their results, we can see that the disc structures are rather different.
\citet{T+14} performed a much higher resolution run ($10^8$ particles).
Unfortunately, however, they only showed snapshots at the early stage of the impact.
In order to test the numerical convergence, we need to follow for the longer time evolutions, at least for $24$ hrs.

In this paper, we present the results of GI simulations with up to $10^8$ particles, focusing on the disc properties.
This paper is organised as follows.
In section \ref{sec:method}, we describe the numerical method.
In section \ref{sec:initial_condition}, we present the initial condition.
In section \ref{sec:results}, we show the results of numerical simulations and also the analysis.
In section \ref{sec:summary}, we summarise this paper.

\section{Numerical Method}\label{sec:method}
In this paper, we employed the standard SPH (for review, see \cite{M92}).
We adopted the Wendland C${}_6$ kernel and the number of neighbour particles for each particle is $\sim 250$.
To handle the shock, we used \citet{M97}'s artificial viscosity.
In order to suppress spurious shear viscosity, we adapted the Balsara switch \citep{B95}.
We solve the self-gravity using the Barnes-Hut tree algorithm \citep{BH86}.
The opening angle for the self-gravity is set to $0.5$.

In order to achieve efficient parallelization, we applied ``\texttt{Framework for Developing Particle Simulator}''\footnote{https://github.com/FDPS/FDPS} (\texttt{FDPS}, \cite{I+15, I+16}).
The central idea of \texttt{FDPS} is to separate a code into two parts: a complex part associated with parallelization and another for the calculation of actual interactions between two particles.
\texttt{FDPS} utilises the tree method to search neighbour lists necessary for the hydrodynamical force and to calculate long-range force quickly.
\texttt{FDPS} can take the full responsibility for the parallelization.

\section{Initial conditions}\label{sec:initial_condition}
In numerical simulations of GI with SPH, each object is represented by a collection of SPH particles.
For both the proto-Earth and the impactor, we assumed that they are differentiated to $30\%$ core (iron) and $70\%$ mantle (granite).
To represent granite and iron, we used the Tillotson equation of state \citep{M89}.
We first place equal-mass particles in 3D Cartesian lattice and then let the particles relax to hydrostatic equilibrium by introducing the damping term \citep{M94}.
The initial specific internal energy for each particle is set to $0.1 G M_\mathrm{\oplus} / R_\mathrm{\oplus}$, where $G$, $M_\mathrm{\oplus}$ and $R_\mathrm{\oplus}$ are the gravitational constant, the mass of the current Earth ($6.0 \times 10^{24}$ kg) and the radius of the current Earth ($6400$ km).
The end time of this process is set to about ten times of the dynamical time.
At the end of this process, the velocity of each particle is of the order of $1 \%$ of the typical impact velocity ($\sim 10$ km/s).
The target-to-impactor mass ratio is assumed to be $0.1$.

After this process, we performed the actual impact simulations.
The initial angular momentum in the system is set to 1.2 times the current angular momentum of the Earth-Moon system ($3.5 \times 10^{34} \, \mathrm{kg \, m^2/s}$).
The impactor velocity at infinity is set to zero so that the impactor takes a parabolic orbit.
Neither objects have initial rotations.

In this paper, we focus on the effect of the resolution to the results.
Hence, we fixed the orbital parameter for all runs and varied the resolutions, namely, the total number of particles $N$.
In this paper, we show the results with $N \sim 3 \times 10^3, 10^4, 3 \times 10^4, 10^5, 3 \times 10^5, 10^6, 3 \times 10^6, 10^7, 3 \times 10^7$ and $10^8$.

\section{Results}\label{sec:results}
\subsection{Predicted moon mass}
In order to compare the results of runs with different numbers of particles quantitively, we use the predicted moon mass (e.g., \cite{I+97, K+00, CA01}) defined as
\begin{equation}
M_\mathrm{M} = 1.9 \frac{L_\mathrm{disc}}{\sqrt{G M_\oplus R_\mathrm{Roche}}} - 1.1 M_\mathrm{disc} - 1.9 M_\mathrm{escape},
\end{equation}
where $L_\mathrm{disc}, R_\mathrm{Roche}, M_\mathrm{disc}$ and $M_\mathrm{escape}$ are the disc angular momentum, the Roche limit ($\sim 2.9 R_\oplus$), disc mass and the mass which will escape during the disc evolution.
We assumed $M_\mathrm{escape}$ to be $0.05 M_\mathrm{disc}$ following \citet{CA01}.
The angular momentum and mass of the disc are given by the sum of the angular momenta and the masses of disc particles.

\begin{figure}[]
	\centering
	\FigureFile(152mm,102mm){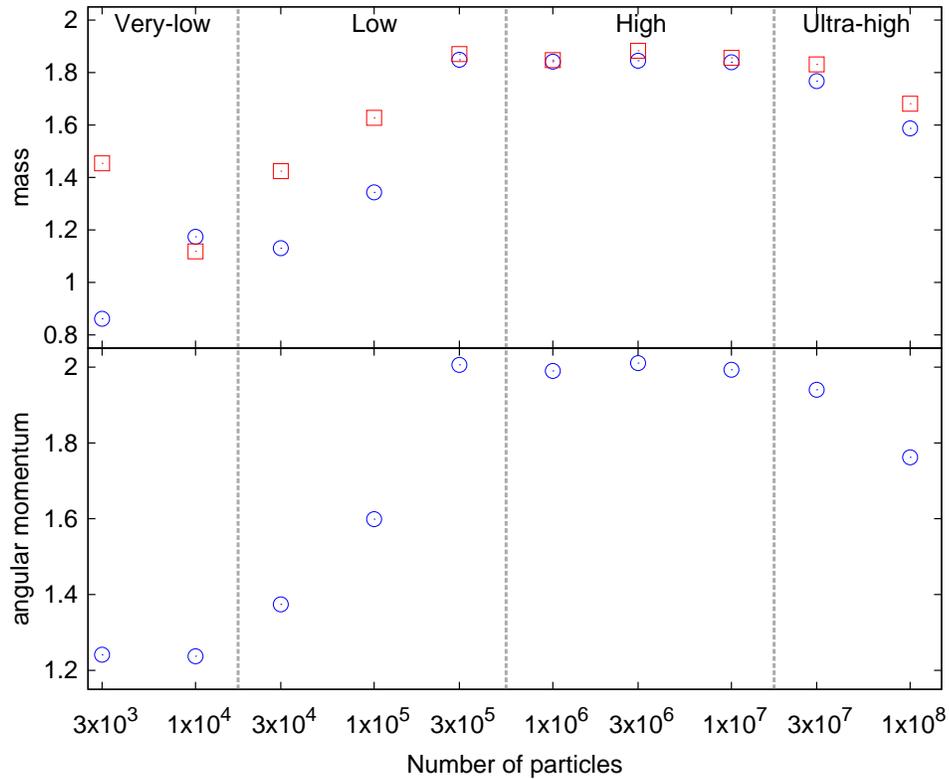}
	\caption{
		Disc properties vs. number of particles.
		The upper panel shows the disc mass (blue circles) and predicted moon mass (red squares) normalised by the current Lunar mass, while the lower panel shows $L_\mathrm{disc} / \sqrt{G M_\oplus R_\mathrm{Roche}}$.
	}
	\label{fig:NvsMM}
\end{figure}

Figure \ref{fig:NvsMM} shows the dependence of the disc quantities on the number of particles.
From this figure we can classify the results into four categories; very-low-resolution runs ($N \leq 10^4$), low-resolution runs ($3 \times 10^4 \leq N \leq 10^5$), high-resolution runs ($3 \times 10^5 \leq N \leq 10^7$) and ultra-high-resolution runs ($3 \times 10^7 \leq N$).
In the very-low-resolution regime, there is no clear dependence of results to the resolution, since $M_\mathrm{M}$, $M_\mathrm{disc}$ and $L_\mathrm{disc}$ all behave differently.
In the low-resolution regime, all show increase as the number of particles is increased, and in the high-resolution regime, it seems all show convergence.
However, this apparent convergence turns out to be an illusion, when we further increase the number of particles and enter the ultra-high-resolution regime, where all values are diverging with no sign of convergence.
In the following, we investigate why this apparent unconvergence occurred in ultra-high resolutions.

Figures \ref{fig:time_evolve_a} and b show the time evolutions of $L_\mathrm{disc} / \sqrt{G M_\oplus R_\mathrm{Roche}}$, $M_\mathrm{disc}$ and $M_\mathrm{M}$ for all runs.
This figure clearly shows that, in the very-low-resolution runs and low-resolution runs, the evolution of the disc depends on the resolution.
In particular, very-low-resolution runs show results quite different from the rest of runs for the whole simulation period [see the top panel of Fig. \ref{fig:time_evolve_a}].
This means that in very-low-resolution runs the number of particles is not enough to resolve the early phase of the simulation, when the impact itself took place ($t < 3$ hrs).
On the other hand, low-resolution runs give similar results for the first $7$ hrs, which means that the early phase is well resolved.
After $7$ hrs, however, the disc properties start to deviate.

On the other hand, high-resolution runs give very similar results.
The disc properties at the end time of the simulations converge to within less than $10\%$ [see the bottom panel of Fig. \ref{fig:time_evolve_b}].
Ultra-high-resolution runs, however, again show the divergence of the disc properties.
If we closely inspect Fig.~\ref{fig:time_evolve_b}, we find that in all runs sudden decreases of $M_\mathrm{disc}$ and $L_\mathrm{disc}$ take place at $t = 8 - 9$ hrs, and that time for $N = 1\times 10^8$ is the latest of all.
Similar, but less clear changes are also visible for runs in Fig. \ref{fig:time_evolve_a}, except for very-low-resolution runs.
Thus, it seems that the mechanism of this sudden decrease might be related to the apparent unconvergence of ultra-high-resolution runs.
In the next subsection, we will investigate what physical mechanism caused these sudden changes and why and how they depend on the resolution.

\subsection{Formation and fallback of clumps}
Figures \ref{fig:full} and \ref{fig:samp} show the snapshots of the results for $t < 10$ hrs for each run and figure \ref{fig:AMcont} shows the angular momentum distribution of high- and ultra-high-resolution runs.
In the case of very-low-resolution runs, no clump is formed in the orbit, while in other runs a clump is formed in the orbit (see the panels from $t = 5$ to $7$ hrs), which eventually falls back to the proto-Earth ($t = 8 - 9$ hrs).
However, the distributions of particles around the clumps depend on the resolution.
In ultra-high-resolution runs, particles with rather high angular momentum ahead of the clump (see the bottom-most row of Fig. \ref{fig:AMcont}).
These particles affect the angular momentum transfer between the clump and the disc.
As a result, the time of the second collision of the clump depends on the resolution.
Once a clump is formed, the disc properties are controlled by the clump.
Since the properties of a clump would depend on the resolution of the disc, runs with different resolutions give different results.

Figure \ref{fig:Cumulative} shows the radial distribution of the cumulative mass for the runs with the number of particles greater than $10^6$.
At $t = 1$ hr, in all runs, the results are similar.
In $t \geq 5$ hrs, however, the differences in the distribution of the mass show up, which come from the angular momentum transfer between the clump and the disc, as we stated above.
In the run with $10^8$ particles, since the clump is more massive than those in the other runs, the disc loses more mass when the clump falls to the proto-Earth.
Thus, the disc mass for the run with $10^8$ particles after the second collision is smaller than those in other runs (see Fig. \ref{fig:time_evolve_b}).

\makeatletter
\renewcommand{\thefigure}{%
\arabic{figure}\alph{enumi}}
\@addtoreset{figure}{enumi}
\makeatother

\setcounter{enumi}{1}
\begin{description}
\item[]
\begin{figure}[]
	\centering
	\FigureFile(152mm,114mm){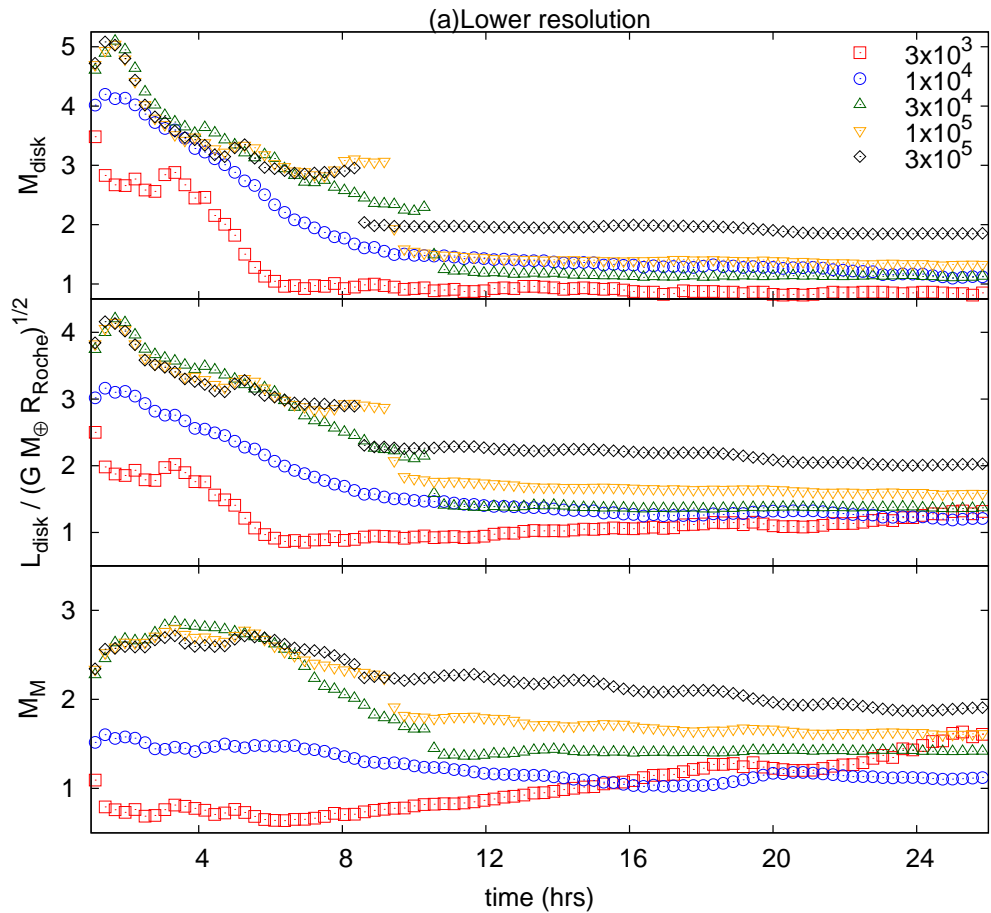}
	\caption{
		Time evolutions of $M_\mathrm{disc}$, $L_\mathrm{disc} / \sqrt{G M_\oplus R_\mathrm{Roche}}$, and $M_\mathrm{M}$ normalised by the current Lunar mass, respectively.
		The meaning of symbols are indicated in the top panel.
		The number of particles is $3\times10^3$ to $3\times10^5$.
	}
	\label{fig:time_evolve_a}
\end{figure}

\addtocounter{figure}{-1}
\setcounter{enumi}{2}
\item[]
\begin{figure}[]
	\centering
	\FigureFile(152mm,114mm){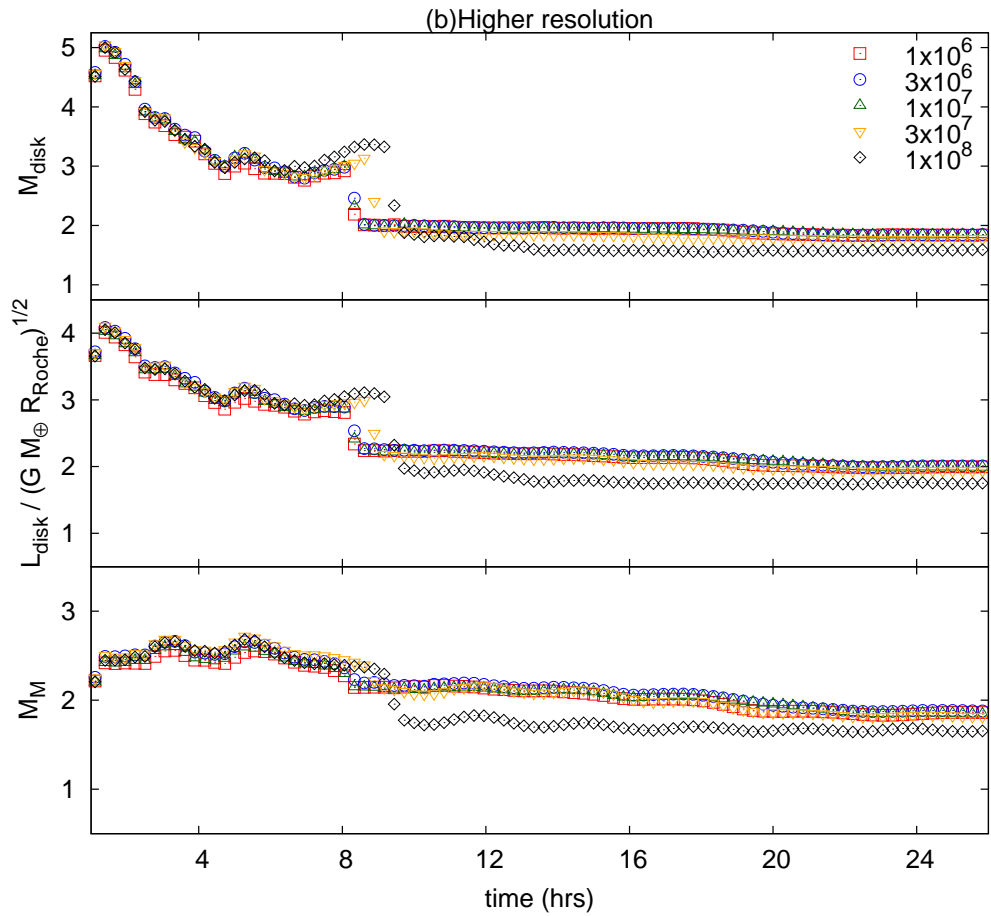}
	\caption{
		Same as Fig. \ref{fig:time_evolve_a}, but for $N=1\times10^6$ to $N=1\times10^8$.
	}
	\label{fig:time_evolve_b}
\end{figure}
\end{description}

\makeatletter
\renewcommand{\thefigure}{%
\arabic{figure}}
\makeatother

\begin{figure}[]
	\centering
	\FigureFile(120mm,120mm){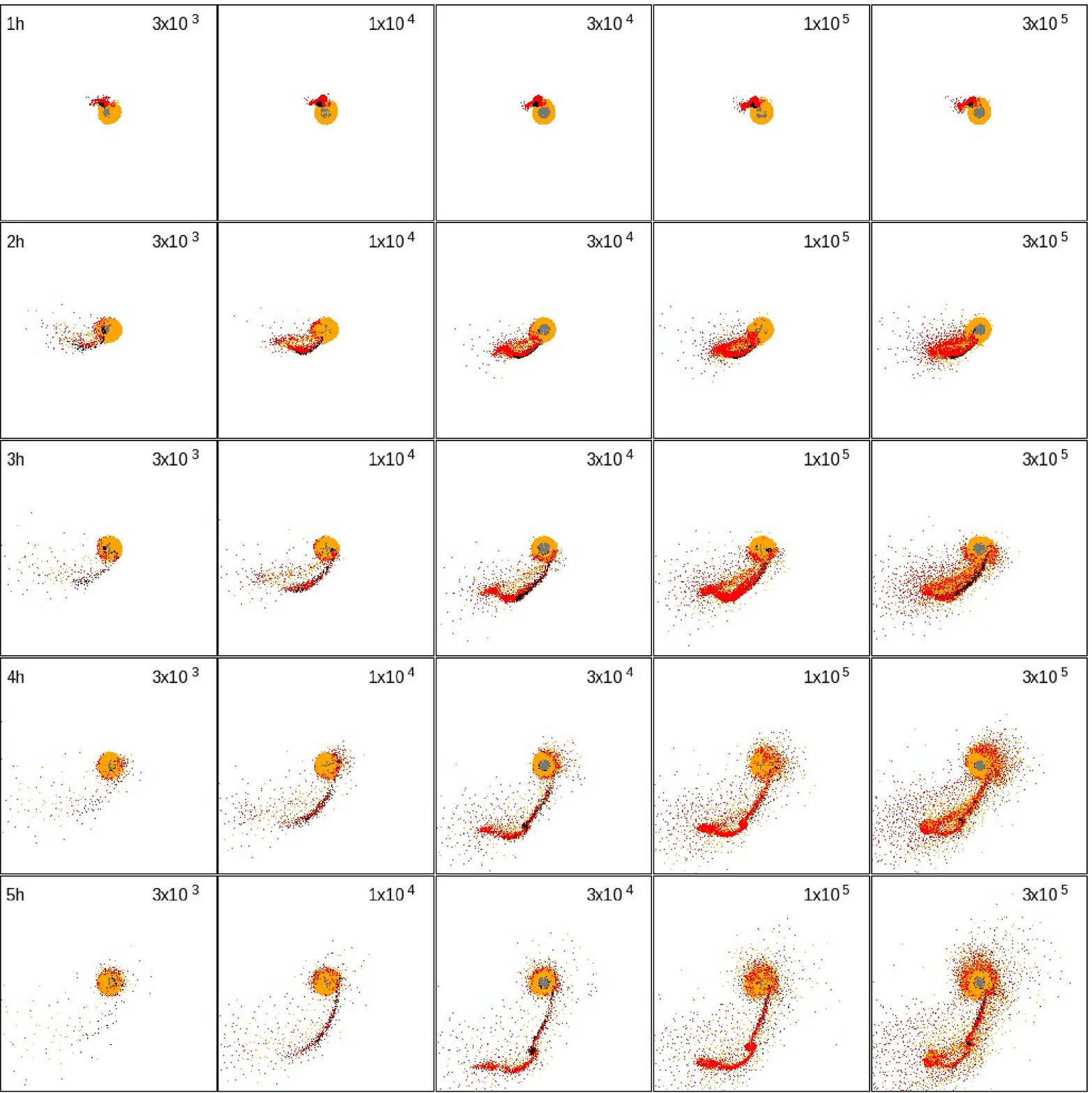}
	\caption{
		Snapshots from runs of GI.
		The snapshot time and number of particles are given in the left-top corner and right-top corner, respectively.
		Orange and red particles indicate mantle particles of the proto-Earth and the impactor, respectively.
		Grey and black particles indicate core particles of the proto-Earth and the impactor, respectively.
		An animation of this figure is available at https://vimeo.com/194156367.
	}
	\label{fig:full}
\end{figure}

\addtocounter{figure}{-1}
\begin{figure}[]
	\centering
	\FigureFile(120mm,120mm){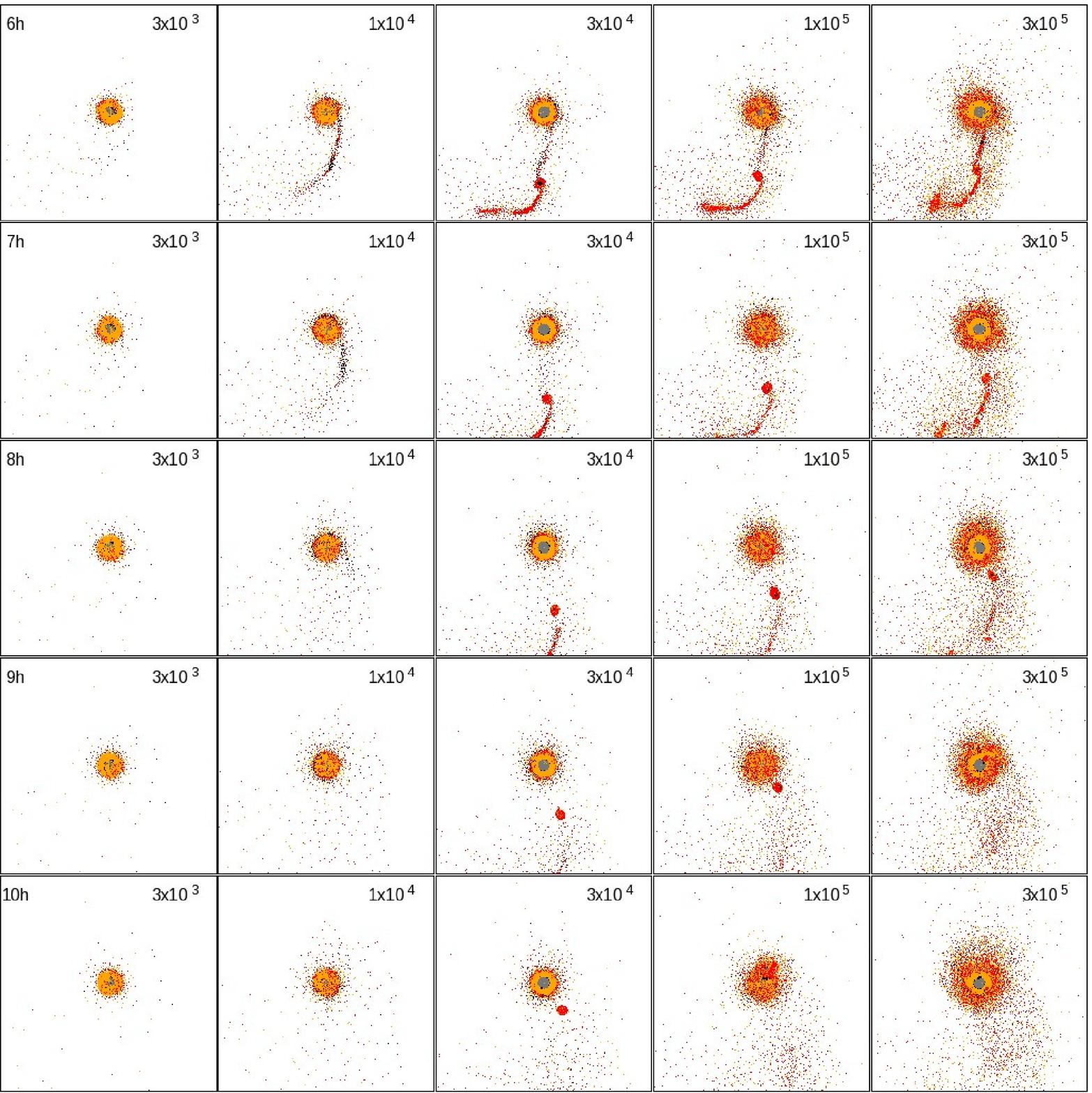}
	\caption{
		Continued.
	}
\end{figure}

\begin{figure}[]
	\centering
	\FigureFile(120mm,120mm){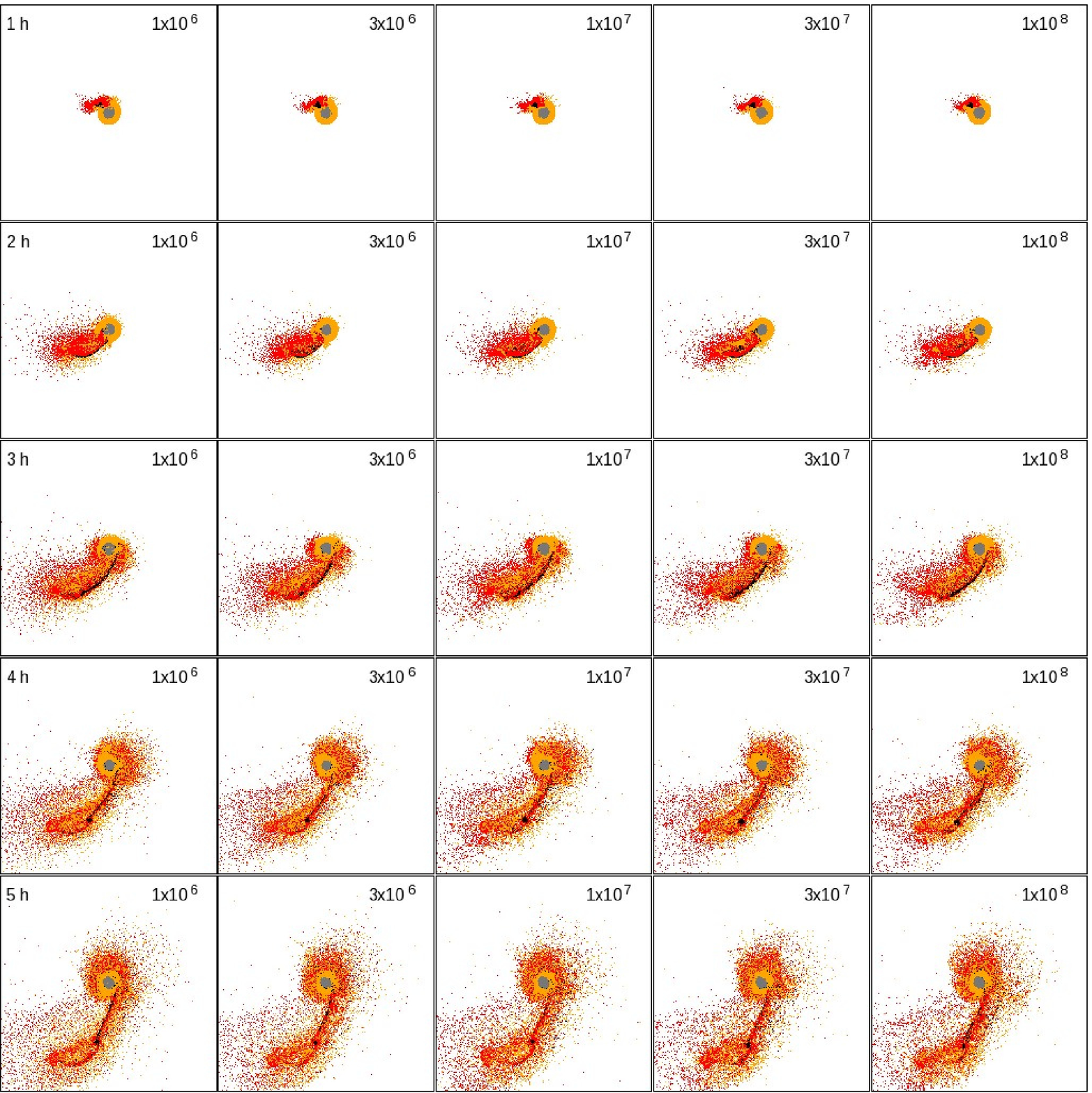}
	\caption{
		Same as Fig. \ref{fig:full}, but reduces the number of plotted particles is reduced to $10^6$.
		Only the results of higher resolution runs are shown.
	}
	\label{fig:samp}
\end{figure}

\addtocounter{figure}{-1}
\begin{figure}[]
	\centering
	\FigureFile(120mm,120mm){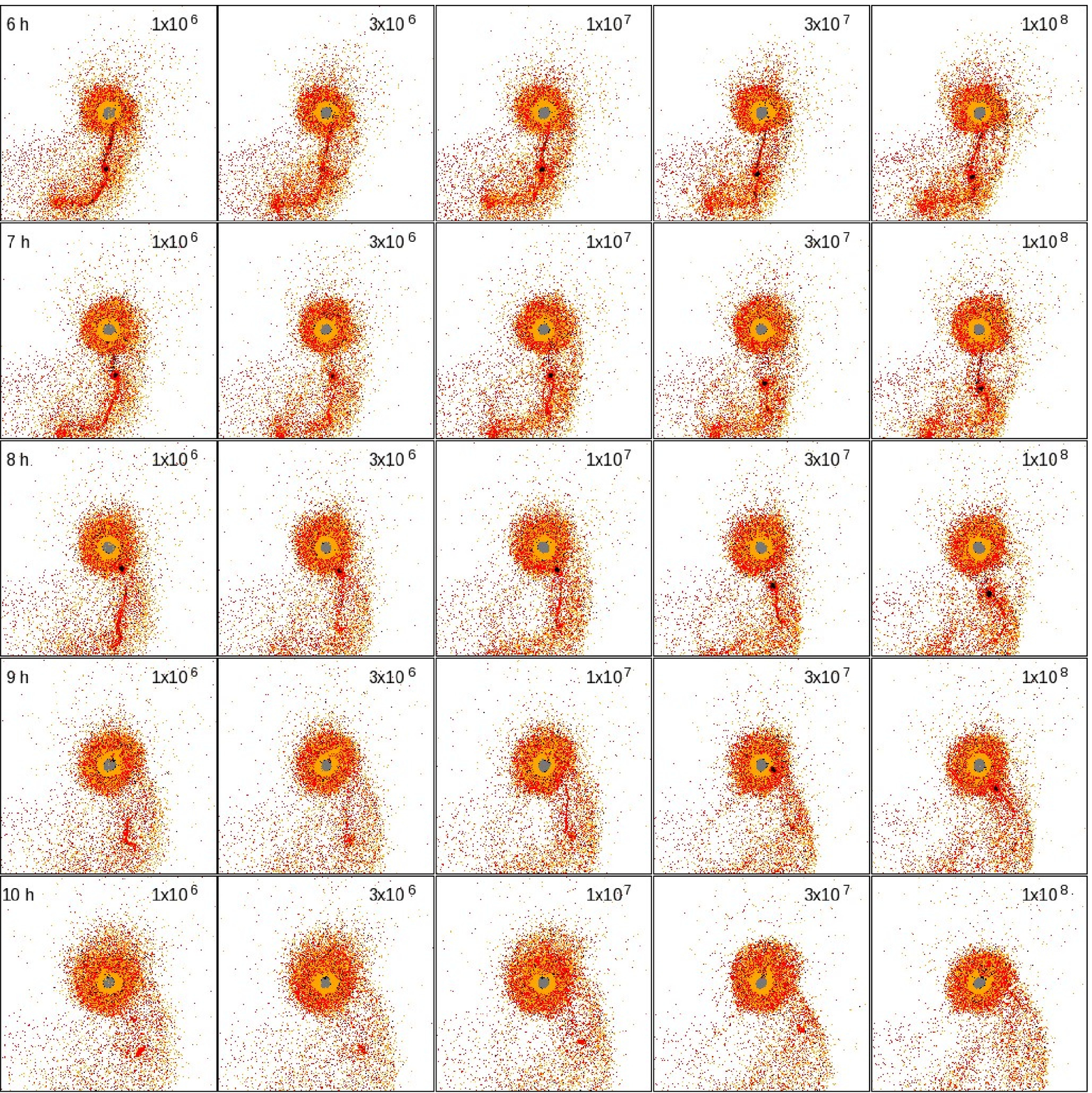}
	\caption{
		Continued.
	}
\end{figure}

\begin{figure}[]
	\centering
	\FigureFile(120mm,120mm){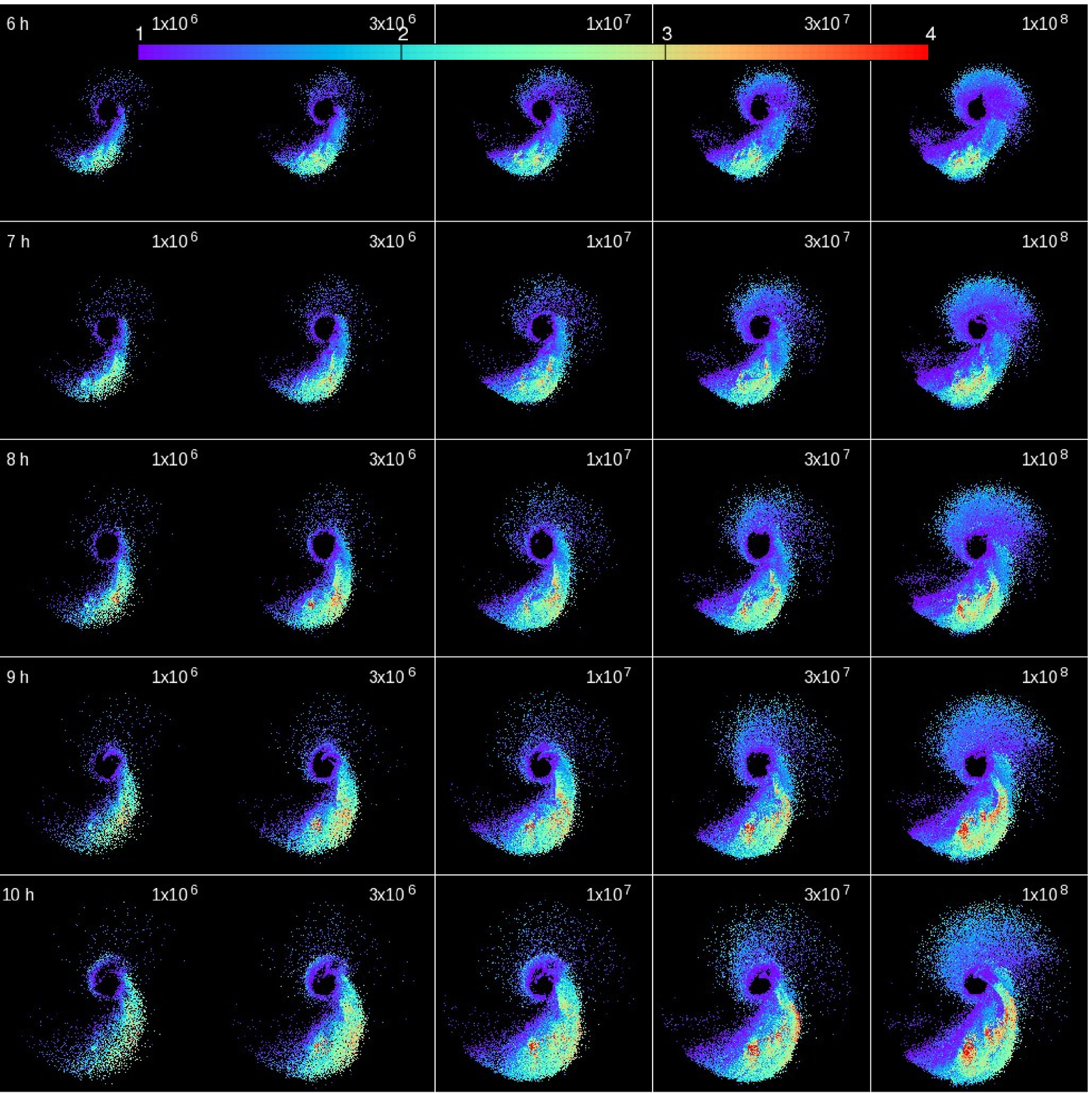}
	\caption{
		Same as Fig. \ref{fig:full}, but shows the distributions of specific angular momentum ($L_\mathrm{disc} / \sqrt{G M_\oplus R_\mathrm{Roche}}$).
		Only the particles within $[-10, 10] \times [-10, 10]$ are shown.
	}
	\label{fig:AMcont}
\end{figure}

\begin{figure}[]
	\centering
	\FigureFile(152mm,203mm){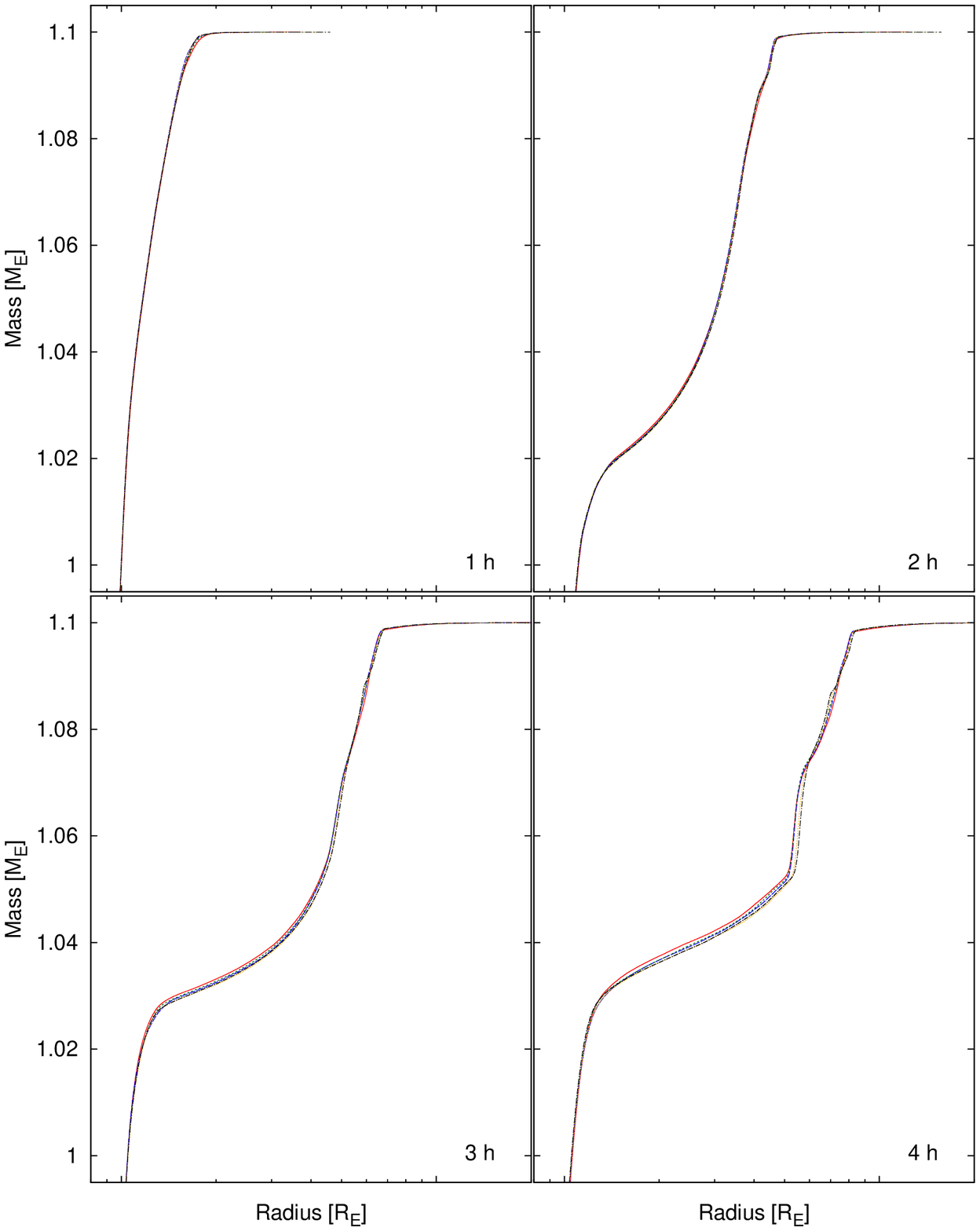}
	\caption{
		Radius vs. the cumulative mass from $t = 1$ to $10$ hrs with an interval of $1$ hr.
		The horizontal axis is shown in log scale.
		The colours for each curve are the same to those of Fig. \ref{fig:time_evolve_b}.
	}
	\label{fig:Cumulative}
\end{figure}

\addtocounter{figure}{-1}
\begin{figure}[]
	\centering
	\FigureFile(152mm,203mm){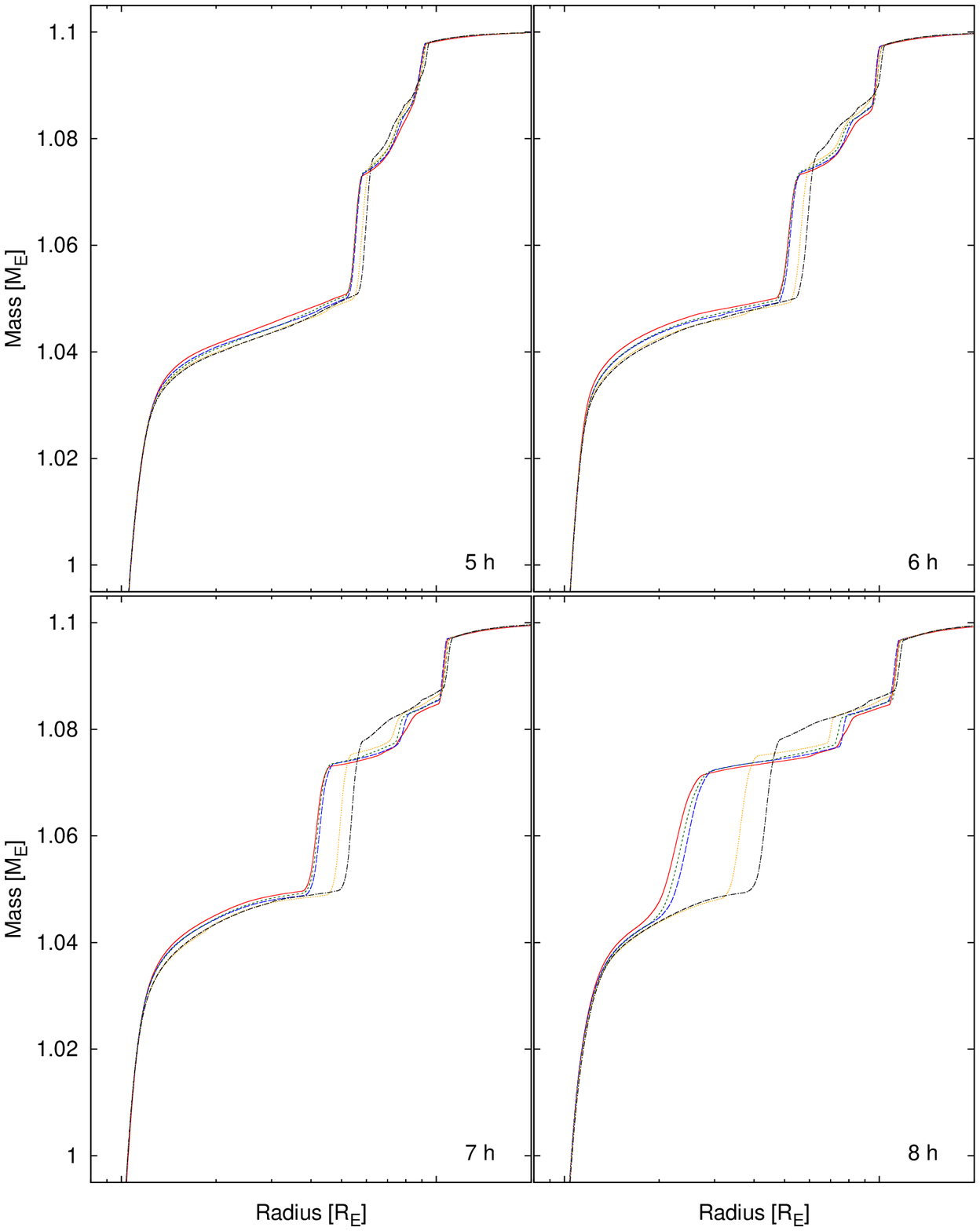}
	\caption{
		Continued.
	}
\end{figure}

\addtocounter{figure}{-1}
\begin{figure}[]
	\centering
	\FigureFile(152mm,203mm){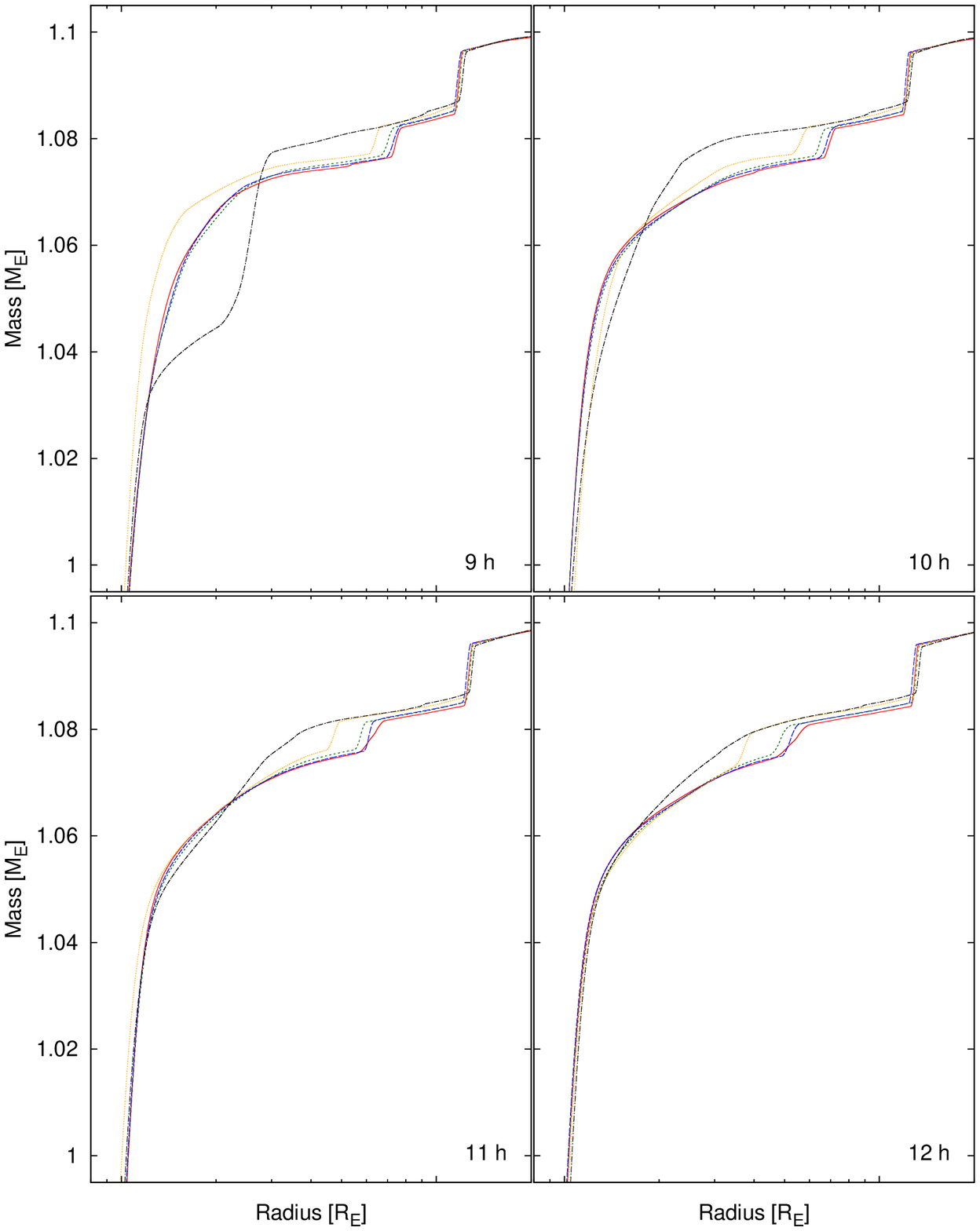}
	\caption{
		Continued.
	}
\end{figure}

\section{Summary}\label{sec:summary}
In order to investigate the effect of the resolution on results of numerical simulations of GI, we carried out runs with a wide range of the number of particles with the standard SPH.
The overall evolutions are similar.
After the first collision of the impactor, large amounts of impactor debris are ejected to the orbit.
The remnant of the impactor then forms a large clump in the orbit and then falls back to the proto-Earth.
This clump changes the detail of the behaviour of those of the properties of the disc, such as the angular momentum distribution.
Thus, a small change in the mass and behaviour of the clump caused a large variation in the disc mass and the predicted moon mass.
This is the reason why the result is apparently not converged for very large number of particles.

Note that our conclusion is consistent with previous work.
\citet{C+13} performed runs with $N = 10^4, 10^5$ and $10^6$ and concluded that the disc properties converged to within $\sim 10\%$.
We, however, found that at the ultra-high-resolution regime, the results change as the number of particles is increased.

The results we show here are certainly not the final ones.
Recently the reliability of the standard SPH was questioned in the context of the treatment of the core-mantle boundary and free surface \citep{H+16}.
In order to test the reliability of numerical simulations of GI, we need to combine scheme comparisons and resolution comparisons.
We plan to perform runs with $10^8$ (or higher) particles with the standard SPH and also DISPH.

\section*{Acknowledgement}
We thank Yutaka Maruyama, Naohito Nakasato and Daisuke Namekata for the cooperations on \texttt{FDPS}.
We also thank Miyuki Tsubouchi for the management of the \texttt{FDPS} developing team.
We are indebted to Takayuki R. Saitoh for his helpful discussion about SPH.
We used computational resources the K computer provided by the RIKEN Advanced Institute for Computational Science through the HPCI System Research project (Project ID:ra000008).

\end{document}